\documentclass[conference]{IEEEtran}
\IEEEoverridecommandlockouts
\usepackage{cite}
\usepackage{amsmath,amssymb,amsfonts,amsthm}
\usepackage{algorithmic}
\usepackage{graphicx}
\usepackage{textcomp}
\usepackage{xcolor}
\usepackage{tikz}
\usepackage{sidecap}
\usepackage{import}
\usepackage{hyperref}
\usepackage{cleveref}
\usepackage{enumitem}
\usepackage{booktabs}
\usepackage{authblk}
\usepackage{fancyhdr}

\def\BibTeX{{\rm B\kern-.05em{\sc i\kern-.025em b}\kern-.08em
    T\kern-.1667em\lower.7ex\hbox{E}\kern-.125emX}}
\newtheorem{lemma}{Lemma}
\newtheorem{theorem}{Theorem}
\newtheorem{corollary}{Corollary}

\sidecaptionvpos{figure}{m}

\usetikzlibrary{calc}
\usetikzlibrary{backgrounds}  
\tikzstyle{proc}=[
    circle,
    minimum size =0.05cm,
    draw=black,
    fill=gray,
    text=white,
    font=\small,
    inner sep=1
]

\definecolor{color1}{HTML}{E9C46A}
\definecolor{color2}{HTML}{2A9D8F}
\definecolor{color3}{HTML}{E76F51}
\definecolor{color4}{HTML}{ffafcc}

\definecolor{color2light}{HTML}{C6E2DF}
\definecolor{color1light}{HTML}{F9EED6}
\definecolor{color3light}{HTML}{F9D6CE}
\definecolor{color4light}{HTML}{FBE9EF}
\definecolor{verylightgray}{HTML}{ECECEC}

\usetikzlibrary{backgrounds}  
\tikzstyle{proc}=[
    circle,
    minimum size =0.05cm,
    draw=black,
    fill=gray,
    text=white,
    font=\small,
    inner sep=1.5
]

\tikzstyle{node} = [
  circle,
  draw=black,
  fill=white,
  text=black
]

\tikzstyle{nodesmall}=[
  node,
  font=\tiny,
  inner sep=2
]

\tikzstyle{child1} = [
  node,
  fill=color2
]

\tikzstyle{child2} = [
  node,
  fill=color3
]
\tikzstyle{child3} = [
  node,
  fill=color4
]

\newcommand{\dist}[0]{\text{dist}}

\author{Yves Baumann}
\author{Tal Ben-Nun}
\author{Maciej Besta}
\author{Lukas Gianinazzi$^{\dag}$\thanks{\dag \ Corresponding author. Email: lukas.gianinazzi@inf.ethz.ch}}
\author{Torsten Hoefler}
\author{Piotr Luczynski$^{\ddag}$\thanks{\ddag \ Corresponding author. Email: pluczynski@student.ethz.ch}}
\affil[]{ETH Zurich, Switzerland \vspace{-1em}}

\hypersetup{
    colorlinks=true,    
    urlcolor=black,      
    linkcolor=black,    
    citecolor=black,     
}
\begin{document}

\title{Low-Depth Spatial Tree Algorithms}

\maketitle

\begingroup
\renewcommand\thefootnote{}\footnote{\copyright 2024 IEEE. Personal use of this material is permitted. Permission from IEEE must be obtained for all other uses, in any current or future media, including reprinting/republishing this material for advertising or promotional purposes, creating new collective works, for resale or redistribution to servers or lists, or reuse of any copyrighted component of this work in other works. The version of record is \url{https://doi.org/10.1109/IPDPS57955.2024.00024}. 
}
\addtocounter{footnote}{-1}
\endgroup

\begin{abstract}
Contemporary accelerator designs exhibit a high degree of spatial localization, wherein two-dimensional physical distance determines communication costs between processing elements. This situation presents considerable algorithmic challenges, particularly when managing sparse data, a pivotal component in progressing data science.
The spatial computer model quantifies communication locality by weighting processor communication costs by distance, introducing a term named \emph{energy}. Moreover, it integrates \emph{depth}, a widely-utilized metric, to promote high parallelism.
We propose and analyze a framework for efficient spatial tree algorithms within the spatial computer model. Our primary method constructs a spatial tree layout that optimizes the locality of the neighbors in the compute grid. 
This approach thereby enables locality-optimized messaging within the tree. Our layout achieves a \emph{polynomial} factor improvement in energy compared to utilizing a PRAM approach. Using this layout, we develop energy-efficient treefix sum and lowest common ancestor algorithms, which are both fundamental building blocks for other graph algorithms. With high probability, our algorithms exhibit near-linear energy and poly-logarithmic depth.
Our contributions augment a growing body of work demonstrating that computations can have both high spatial locality \emph{and} low depth. Moreover, our work constitutes an advancement in the spatial layout of irregular and sparse computations.
\end{abstract}


\section{Introduction}\label{sec:introduction}

Spatial computing architectures have emerged as significant platforms in optimizing energy use and enhancing throughput, often outperforming traditional CPUs and GPUs in specific applications. Notable among these architectures are the Cerebras Wafer-Scale Engine (WSE) \cite{Cerebras,DBLP:journals/corr/abs-2209-13768,DBLP:conf/sc/RockiESSMKPDS020,Trifan2022CerebrasCov,DBLP:journals/corr/abs-2304-03208} and Coarse-Grained Reconfigurable Arrays~\cite{DBLP:conf/asap/ChinSRZKHA17, DBLP:conf/fpga/GaideGRB19,DBLP:conf/fpga/Vissers19, DBLP:conf/fpga/SwarbrickGAGA19, DBLP:journals/access/PodobasSM20}. These platforms employ highly localized memory systems, wherein each processor is allocated its own segment of fast memory, and feature a communication network where the cost between processors is dictated by their relative proximity. This paradigm proves particularly advantageous in applications such as matrix computations~\cite{Cerebras2023GordonBell}  and stencils~\cite{DBLP:conf/sc/RockiESSMKPDS020}. Yet, it also introduces programming and algorithmic challenges, especially when computing on sparse data---a widespread and growing challenge in domains like machine learning~\cite{DBLP:journals/air/Kotsiantis13,DBLP:journals/ml/Breiman01} and deep learning~\cite{Torsten2021, ben2019modular, besta2022parallel, zhang2020deep}.

The emerging focus on sparsity and its resulting irregular access patterns underscore a critical challenge for spatial architectures, prompting the exploration and development of innovative algorithmic approaches to address these complexities. This work introduces the first \emph{spatial tree algorithms}, presenting a noteworthy advancement in optimizing irregular access patterns for spatial architectures. 
Trees are fundamental graph structures, and tree algorithms often underpin more complex graph algorithms~\cite{KargerMinCut}. They offer wide applicability and utility across various scientific and technological domains.

The optimization of tree layouts has direct applications in diverse fields such as machine learning and computational biology. In the former, models like decision trees~\cite{DBLP:journals/air/Kotsiantis13} and random forests~\cite{DBLP:journals/ml/Breiman01} can realize enhanced performance through spatial locality. Meanwhile, in computational biology, the study of phylogenetic trees~\cite{piel2009, vos2012} provides vital insights into evolutionary pathways~\cite{Pennisi2003} by extensively analyzing tree structures. This highlights the broad relevance and widespread applicability of optimized tree layouts and algorithms in various scientific domains.

\subsection{Methodology}

We analyze our results in a model that captures the main characteristics of \emph{spatial} architectures~\cite{SpatialModel}. 
The model abstracts certain lower-level details such as the interconnection network, but emphasizes the main drivers of performance. Let us consider the Cerebras WSE-2 as an example~\cite{Cerebras}. It has 850,000 cores with around 50KB of fast local memory per core. Each processor can send and receive a 32-bit message per cycle. The time to reach the destination has an initial latency of 2 cycles and is then proportional to the distance on the chip. Hence, it is important to minimize (1) the distance traveled and (2) the number of hops in the communication chains. The \emph{spatial computer model}~\cite{SpatialModel} encourages these qualities and enables productive algorithm design by focusing on the main performance factors.

The model considers a two-dimensional grid of processors. Each processor \(p_{i, j}\) has a position \((i, j)\) and a local memory containing a \emph{constant} number of words. In each round, a processor can send and receive a constant number of messages and then perform a constant number of arithmetic operations on its local memory. The \emph{energy} cost of a processor \(p_{i, j}\) sending a message to a processor \(p_{x, y}\) is equal to \(|x - i| + |y - j|\), i.e., the Manhattan distance. The energy of an algorithm is the sum of the energy of all its messages. Essentially, the energy is the distance-weighted communication volume. The \emph{depth} of a spatial computation is the longest chain of messages depending on each other.

Reducing the \emph{energy} leads to more spatially local algorithms that minimize the distance traveled; this, in turn, enhances the efficiency and sustainability of large-scale applications. By minimizing the \emph{depth} of the model, shorter communication chains can be achieved, enabling algorithms to effectively leverage the vast number of compute cores available in modern spatial architectures.  
Although the model is formulated for processors with asymptotically constant-size memory, its results are applicable in settings where processors have larger memories of size $M$~\cite{SpatialModel}. Hence, by focusing on bounded memory, we obtain results for general memory sizes.

\subsection{Limitations of State-of-the-Art}

None of the existing approaches capture the unique requirements of having high spatial locality and low depth:

\subsubsection*{PRAM} There is a rich literature of low-depth and work-optimal algorithms on trees~\cite{MiRe85, SchieberLCA1988, DBLP:conf/icci/LinO91, LinLCA1992, DBLP:journals/siamcomp/BerkmanV93}. However, the PRAM model does not have any notion of spatial locality and consequently its algorithms exhibit suboptimal energy. However, we will show how to adapt principles of parallel algorithm design. When combined with our specialized data layouts, this adaptation achieves low-depth \emph{and} low-energy. 

\subsubsection*{Parallel External Memory}
Arge et al.~\cite{pem_lca} presented parallel external memory (PEM) algorithms for list ranking, expression tree evaluation, and lowest common ancestors. In PEM~\cite{DBLP:conf/spaa/DehneDH97}, it is crucial to subdivide the working set into chunks that fit into the processors' local fast memory. This issue is even more pronounced in the spatial setting, as the processors have only constant-sized memory. In contrast to the PEM setting, obtaining near-linear energy bounds precludes sorting the data.

\subsubsection*{CGM} Dehne et al.~\cite{bsp_lca} considered lowest common ancestors and expression tree evaluation in the Coarse-Grained Multicomputer model (CGM). In CGM~\cite{DBLP:journals/ppl/ChanD99, DBLP:journals/ijcga/DehneFR96}, the number of processors is smaller than the input size by a polynomial factor. Hence, CGM work focuses on reducing communication at the expense of parallelism. In contrast, the spatial computer considers the setting where the processor count is on a similar order of magnitude as the input size, thereby demanding maximized parallelism.

\subsubsection*{FCN} Leighton~\cite{Leighton1991IntroductionTP} proposed graph algorithms for fixed-connection networks (FCN), where the communication network has a fixed topology, for example, a mesh of trees~\cite{DBLP:conf/wg/Leighton83}.  For dense graphs, their approach embeds the graph in a fixed network of \emph{quadratic size} in the number of nodes. For sparse graphs, their approach relies on PRAM simulation. 
In contrast, the spatial computer is network-oblivious in that the energy and depth terms are meaningful for a variety of topologies~\cite{SpatialModel}. Moreover, we avoid costly PRAM simulations.

\subsubsection*{VLSI Complexity} The problem of embedding graphs onto a two-dimensional plane has received much attention in VLSI Complexity~\cite{Leighton2003ComplexityII, DBLP:conf/stoc/LiptonS81}. Similar to our setting, reducing wire lengths is a central goal in VLSI. However, our model differs significantly in both its goals and approaches. VLSI concentrates on hardware wiring, which is inherently static, focusing on the physical layout of connections. Our model is tailored for algorithm design, emphasizing dynamic message passing rather than static wire configuration. This fundamental difference in objectives leads to distinct methodologies. For instance, unlike VLSI, our model does not prioritize crossing-free wire placement. Instead, our emphasis is on optimizing the efficiency of message passing. Consequently, we use proxy metrics such as energy and depth to evaluate the routing of messages. These metrics are versatile, applicable across a variety of communication networks and routing policies. 

\subsection{Key Insights and Contributions}

Tree algorithms are nontrivial to make spatially local because traditional algorithms exhibit irregular access patterns. Consider, for instance, a fundamental tree kernel that sends a message from each vertex to its children. Although this \emph{local messaging} kernel has constant depth, its energy depends on the layout of the tree on the processor grid. The average distance between neighbors varies dramatically based on vertex and edge placements: A suboptimal layout can yield a communication distance between vertices that deviates polynomially from the optimum. 

We present a locality-optimized tree layout to obtain an efficient message-passing kernel on trees. 
The layout has two main steps: First, we map the tree onto a linear order based on the sizes of the subtrees using an order we call \emph{light-first}. Second, the linear order is mapped to the 2D grid with a space-filling curve~\cite{SpaceFillingBook}. Our analysis shows that this layout leads to a message-passing kernel with \emph{linear energy}, which is optimal up to constant factors. 
This is achieved while preserving the low depth characteristic of the local messaging kernel. For this result, we bound the distances in several space-filling curves. Trees of unbounded degree need special care because of the limitations imposed by having $O(1)$ memory per processor. Nevertheless, we attain the same energy bounds on general trees for the following useful communication patterns: when children uniformly receive a message from a parent and when a parent receives reduced messages from its children.

We demonstrate how to utilize this locality-optimal layout to implement treefix sums and lowest common ancestors (LCA), two pivotal tree operations, with near-linear energy and poly-logarithmic depth. These operations are subroutines for other graph algorithms, such as the computation of minimum cuts~\cite{KargerMinCut}. To formulate efficient algorithms within the local messaging framework, specific adaptations are necessary. For treefix sums, this requires adapting low-depth techniques to the spatial setting. For LCA, this requires the design of a new approach based on a cover of the tree with carefully chosen subtrees. This cover uses our efficient treefix sum algorithm.  We present randomized (Las Vegas) algorithms for treefix sum and batched lowest common ancestors that take \(O(n\log n)\) energy and  $O(\log ^2 n)$  depth with high probability on a tree with $n$ vertices. If the tree has bounded degree, the depth of the treefix sum algorithm is $O(\log n)$. In comparison, the simulation of a work-optimal PRAM algorithm would take $\Theta(n^{\frac{3}{2}})$ energy and $O(\log ^4 n)$ depth.

\subsection{Limitations of the Proposed Approach}

To achieve the best performance, we must store the trees in our efficient layout before executing the tree algorithms. We can mitigate this limitation and amortize the layout costs when using the same tree across multiple iterations, a common scenario in machine learning and phylogenetic applications.
\section{Background}\label{sec:background}

\begin{table}[]
    \centering
    \caption{Tree and Cost Model Notation}
    \renewcommand{\arraystretch}{1.2}
    \label{tab:notation}
    \begin{tabular}{ll}
    \toprule
    Symbol & Description \\
    \midrule
    $T, \hat T, T' \dotsc $ & A rooted tree \\
    $n$ & Number of vertices in $T$ = number of processors \\
    deg$(v)$ & degree of vertex $v$ \\
    $\Delta$  & Maximum vertex degree in the tree $T$\\
    $s(v)$ & number of descendants of $v$ \\
    $E(n), E_d, \dotsc $ & Energy, i.e., total Manhatten distance traversed \\
    \bottomrule
    \end{tabular}
\end{table}

We proceed to give the background needed for our technical discussion, including on our model of computation and space-filing curves, which are a crucial tool in our constructions.

\subsection{Model of Computation}

In spatial architectures, the physical distance has a direct impact on the cost of communication~\cite{iff2023hexamesh, iff2023rapidchiplet}. Longer distances increase latency, indicate potential congestion, and the active energy is proportional to total distance traveled. Because of manufacturing constraints, many commercially available chips, including the Cerebras WS-2, use mesh-like interconnects, exacerbating these issues~\cite{DBLP:conf/sc/JacquelinAM22}.

The spatial computer model provides a suitable abstraction for such architectures~\cite{SpatialModel}.
It considers a \(\sqrt{n} \times \sqrt{n}\) grid of processors with constant-sized memory each. We consider a computation as a directed acyclic graph, where the vertices correspond to computation at a given processor in the grid and the edges correspond to communication. Each edge has a weight corresponding to the Manhattan distance between the processors of the computations it connects.  Processors may compute independently, but a computation waits for all incoming messages on which it depends. The largest number of messages in a chain of dependent messages is the \textbf{depth} of the computation. The depth provides a measure of how often the computation switches between communication and computation steps, which can be costly. A low-depth algorithm can be scheduled in a small number of parallel steps~\cite{DBLP:journals/jacm/Brent74}. The total weight of the communicating edges is the \textbf{energy} of the computation. The energy provides an estimate on the cost of routing the messages, as a larger energy corresponds to sending messages for more network hops. Note that the energy is bounded by the work, i.e., the total number of operations, because the constant-sized memory implies that an algorithm performs $O(1)$ computations between sending two messages. 

We employ the following foundational spatial algorithms. 
The \textbf{Broadcast} of a scalar value from one processor to all other processors in the grid takes \(O(n)\) energy and \(O(\log n)\) depth. A \textbf{reduce} operation computes the sum of a set of $n$ values and takes \(O(n)\) energy and \(O(\log n)\) depth. An \emph{all-reduce}, which is a reduce followed by a broadcast, has the same energy and depth bounds \cite{SpatialModel}. Parallel \textbf{prefix sum} takes $O(n)$ energy and poly-logarithmic depth.
\textbf{Sorting} $n$ numbers takes $\Theta(n^{\frac{3}{2}})$ energy and poly-logarithmic depth, which matches the $\Omega(n^{\frac{3}{2}})$ energy lower bound of a global permutation on a $ \sqrt{n}\times\sqrt{n}$ grid~\cite{SpatialModel}.

The spatial computer can perform a \textbf{PRAM simulation} of any shared-memory parallel algorithm. If an algorithm uses $p$ processors, $m$ memory cells, and $T_p$ time steps, it takes $O( p (\sqrt{p}+\sqrt{m}) T_p)$  energy with poly-logarithmic depth overhead. While typically resulting in sub-optimal algorithms, this simulation easily provides upper bounds. 

A \textbf{Las Vegas} algorithm always produces the correct result with a probabilistic guarantee on its costs. For all our Las Vegas algorithms, the bounds hold with high probability. This means that given any constant \(c > 0\), the probability that the bound holds is greater than 1-\(\frac{1}{n^c}\).

\subsection{Space-Filling Curves}

A key element used in our tree layout schemes are \emph{space-filling curves} \cite{SpaceFillingBook}. A discrete space-filling curve maps a subset of the natural numbers onto a subset of the 2D grid. 

We define the \textbf{Hilbert curve} \cite{HilbertCurve,SpaceFillingBook} of \(k\)-th order for a grid of size \(4^\frac{k}{2}\times 4^\frac{k}{2}\) by dividing it into four subgrids, each with a Hilbert curve of \((k- 1)\)-th order, flipping the two lower curves across the diagonals and then connecting the subgrids. 
A Hilbert curve of 0-th order is a \(1 \times 1\) square. If a curve has order \(k\), we assume it is defined on a \(4^\frac{k}{2}\times 4^\frac{k}{2}\) subgrid. See Figure \ref{fig:tree_layout} (right) for an example.

The \textbf{Z-order curve} \cite{SpaceFillingBook,HaverkortLocalityBounding} is defined by dividing the grid into four quadrants. Visit the four quadrants recursively in the following order: upper left; upper right; lower left; lower right. See Figure \ref{fig:z_order} for an example. Note that the Z-order curve is not distance-bound. However, we will show that it provides the same spatial locality properties as the other curves.

\subsection{Trees}

Throughout, we consider a rooted tree $T=(V, E)$ with $n=|V|$ vertices. The degree of a vertex $v$ counting its parent and children is deg$(v)$ and the maximum degree in the tree $T$ is $\Delta$. The descendants of a vertex $v$ contains $v$, its children, and recursively the descendants of its children. A vertex $u$ is an ancestor of a vertex $v$ if $v$ is a descendant of $u$. \Cref{tab:notation} provides a summary of our notation.

\section{Spatial Tree Layouts}

We show how to embed a tree into the two-dimensional grid so neighbors can communicate with \emph{constant} average energy. This embedding is the key tool for our tree algorithms in Sections \ref{sec:treefix} and \ref{sec:lca}.
The main idea is as follows: First, compute a linear order of the vertices of the tree. Second, embed the linear order using a space-filling curve.

The challenge lies in finding the correct linear order and proving that the linear order indeed results in a low-energy layout. Na\"ive solutions, such as using breadth-first order or depth-first order, do not yield the desired results. In particular, a perfect binary tree will have a breadth-first layout where the average distance between neighbors is $\Omega(\sqrt{n})$. For depth-first order, a tree formed by adding an additional vertex as a child of each vertex in a path graph provides similarly poor results. Hence, it is crucial to optimize the tree layout.

We will first consider trees of bounded degree and then generalize the results to arbitrary degrees in \Cref{sec:unbounded}.
\subsection{Light-First Order}\label{storing algorithm}

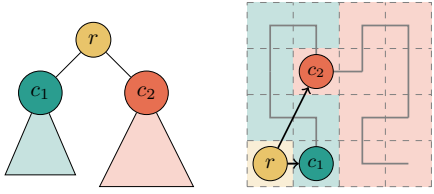
\begin{figure}
\begin{center}
  \vspace{-1em}
  \resizebox{0.35\linewidth}{!}{
\begin{tikzpicture}[step=0.75cm, x=0.75cm, y=0.75cm, node distance=12mm]

    \node[node,fill=color1] (r) {$r$};
    
    \node[child1] (c1) [below left of=r] {$c_1$};
    \node[child2] (c2) [below right of=r] {$c_2$};
    
    \draw (r) -- (c1);
    \draw (r) -- (c2);
    
    \draw[draw=none] (2.25,0.1) -- (2.75,0.1);
    \begin{scope}[on background layer]
      \draw[fill=color2light] (c1) -- +(-0.75,-1.75) -- +(0.75,-1.75) -- (c1);
      \draw[fill=color3light] (c2) -- +(-1,-2) -- +(1,-2) -- (c2);
    \end{scope}
    
\end{tikzpicture}
}
\resizebox{0.3\linewidth}{!}{
\begin{tikzpicture}[step=0.75cm, x=0.75cm, y=0.75cm]

    \fill[fill=color1light] (0,0) rectangle (1,1);
  
    \fill[fill=color2light,] (1,0) rectangle (2,1);
    \fill[fill=color2light,] (0,1) rectangle (2,2);
    \fill[fill=color2light,] (0,2) rectangle (1,4);
    \fill[fill=color2light,] (1,4) rectangle (2,3);
  
    \fill[fill=color3light] (1,3) rectangle (2,2);
    \fill[fill=color3light] (2,4) rectangle (4,0);

    \draw[help lines, dashed] grid +(4,4);

    \draw[thick, color = gray] (0.5,0.5) -- ++(1, 0) -- ++(0,1) -- ++(-1,0) -- ++(0,1);
    \draw[thick, color = gray] (0.5,2.5) -- ++(0,1) -- ++(1,0) -- ++(0,-1) -- ++(1,0);
    \draw[thick, color = gray] (2.5,2.5) -- ++(0,1) -- ++(1,0) -- ++(0,-1) -- ++(0,-1);
    \draw[thick, color = gray] (3.5,1.5) -- ++(-1,0) -- ++(0,-1) -- ++(1,0);
    
    \node[node,fill=color1] (r) at (0.5,0.5){$r$};
    \node[child1,font=\small,inner sep=2.25] (c1) at (1.5,0.5){$c_1$};
    \node[child2,font=\small,inner sep=2.25] (c2) at (1.5,2.5){$c_2$};
    \draw[->,thick] (r) -- (c1);
    \draw[->,thick] (r) -- (c2);
  
\end{tikzpicture}
}
  \vspace{-2.5em}
  \caption{Part of a tree stored in Hilbert-light-first order. The smaller subtree is stored first, then the larger subtree follows. Both subtrees are stored similarly recursively. Mapping this linear order onto the Hilbert curve yields an energy-efficient two-dimensional layout.}
  \label{fig:tree_layout}
  \vspace{-1em}
\end{center}
\end{figure}

We define a tree layout, called \textbf{light-first order} (or smallest-first order~\cite{Gianinazzi2024Arrow}), and show that trees stored in this order allow for energy-efficient messaging. 
We can define light-first order for any space-filling curve. Let \(v\) be some vertex with children \(\{c_1, ..., c_{\deg(v)}\}\). Let \(s(c_i)\) be the size of the subtree rooted at the vertex \(c_i\). Let us now assume that a vertex \(v\) is stored in a processor \(p_v\) and that the children are indexed in increasing order of their subtree size. We say that a vertex \(v\) has a \emph{neighborhood} stored in \emph{light-first order} if each \(c_i\) is stored in \((1 + p_v + \sum_{j=1}^{i - 1}s(c_j))\)-th position according to the space-filling curve. Note that a vertex with no children has a neighborhood in light-first order by default. We say that a tree \(T\) is in \emph{light-first order} if every vertex \(v\) in \(T\) has a neighborhood stored in light-first order. We can say \emph{K-light-first order} to specify that it is defined for the space-filling curve K. See Figure \ref{fig:tree_layout} for an example of a tree stored in Hilbert-light-first order.

\subsection{Distance-Bound Curves}
\setlength{\abovedisplayskip}{0.5em}
  
Consider a tree $T$ with a degree bounded by some constant \(\Delta\). Let that tree be stored in light-first order for some space-filling curve. 
For a space-filling curve, we say that a processor is \(i\)-th if its location on the grid corresponds to the \(i\)-th element in that curve's order. 
We say that a curve is \textbf{distance-bound} if for all natural numbers $i$ and $j$, sending a message from the \(i\)-th processor to the \((i + j)\)-th processor takes \(O(\sqrt{j})\) energy.
We will get different constant factors depending on the type of distance-bound curve. That is, sending the message takes \(c \sqrt{j} + o(\sqrt{j})\) energy for some constant $\alpha$. Examples include the Hilbert curve with \(\alpha = 3\)~\cite{niedermeier_manhattan-distance_1996}, the Peano curve with \(\alpha = \sqrt{10 + 2/3}\)~\cite{SpaceFillingBook}, the \(\beta\Omega\) curve with \(\alpha = 3\)~\cite{SpaceFillingBook,HaverkortLocalityBounding}, or the H-index where \(\alpha = 2\sqrt{2}\)~\cite{SpaceFillingBook,Niedermeier.etal02}. There are also other space-filling curves for which the constant \(\alpha\) has been proven~\cite{SpaceFillingBook,Niedermeier.etal02,niedermeier_manhattan-distance_1996}. The definition also applies to space-filling curves not defined for a square grid, e.g., the Gosper Flowsnake. However, in a practical sense, non-square curves may be rather inefficient.

We say that a light-first order is \textbf{energy-bound} if, for any tree stored in this order, the total energy of each vertex sending a message to all its children is \(O(n)\).
In this section, we prove that any light-first order for a distance-bound space-filling curve is also \emph{energy-bound}.

Let \(\dist(i, j)\) be the energy cost of sending a message from the \(i\)-th position to the \(j\)-th position. If the curve is distance-bound, \(\dist(i, j) \leq c \cdot \sqrt{|j - i|}\) for some \(c \in O(1)\). 
For simplicity, we assume that all vertices have exactly $\Delta$ children. If a vertex has less than $\Delta$ children, we can introduce empty subtrees to make it so.

%
%
Let \(E(n)\) be the energy cost of sending the messages in a subtree of size \(n\). It is bounded as follows:
\begin{lemma}
  \(E(n) \leq (\sum_{i = 1}^\Delta E(s({c_i})) + (\Delta - i) \cdot c\sqrt{s({c_i})}) + \Delta \cdot c\sqrt{2}\).
\end{lemma}
\begin{IEEEproof}
We can express $E(n)$ recursively:
\[
E(n) \leq \sum_{i = 1}^{\Delta} E(s({c_i})) + \dist\left(p_r, p_r + 1 + \left(\sum_{j = 1}^{i - 1}s({c_j})\right)\right)  \enspace .
\]
 From the definition of distance-bound, \(\dist(p_r, p_r + 1 + \sum_{j = 1}^{i - 1}s({c_j})) \leq c \sqrt{2 + \sum_{j = 1}^{i - 1}s({c_j})}\), which implies the result.

\end{IEEEproof}
We need the following lemma to prove the main theorem.
\begin{lemma} \label{minimized}
  If \(s({c_i}) \leq s({c_j}), \forall j \geq i\) and \(\sum_{i = 1}^ds({c_i}) = n\), the equation \(\sum_{i = 1}^\Delta (\Delta + i) \cdot c\sqrt{s({c_i})}\) is minimized for \(s({c_d}) = n\).
\end{lemma}
For the proof of \Cref{minimized}, see Appendix \ref{sec:appendix}.

\begin{theorem}\label{total message}
  Light-first order defined on a distance-bound space-filling curve is energy-bound.
\end{theorem}

\begin{IEEEproof}
We prove by strong induction on the number $n$ of elements in the tree that 
\begin{align*}
  E(n) \leq \Delta \cdot 8c \cdot n - \Delta \cdot 2c \sqrt{n} - 2c \sqrt{2}, \forall n \in \mathbb{N} \enspace .
\end{align*}
The base case for \(n = 1\) follows from \(c \geq 1 \). 
The inductive step goes as follows: 
  \begin{displaymath} \label{eq1}
    \begin{split}
      &E(n + 1) \leq \Delta \cdot c\sqrt{2} + \sum_{i = 1}^\Delta E(s({c_i})) + (\Delta - i) \cdot c\sqrt{s({c_i})} \\
       &\leq^{\text{(I.H.)}}  \Delta \cdot c\sqrt{2} + \sum_{i = 1}^\Delta \Delta \cdot 8c \cdot s({c_i}) - \Delta \cdot 2c \sqrt{s({c_i})} \\ 
      & \quad \quad  - 2c \sqrt{2} + (\Delta - i) \cdot c\sqrt{s({c_i})}\\
      &\leq \Delta \cdot 8c \cdot n - \Delta \cdot c\sqrt{2} - \sum_{i = 1}^\Delta (\Delta + i) \cdot c\sqrt{s({c_i})} \enspace .
    \end{split}
  \end{displaymath}
By Lemma \ref{minimized}, the expression is minimized for \(s({c_\Delta}) = n\). Hence, 
\begin{align*}
E(n + 1) 
&\leq \Delta \cdot 8c (n + 1) - \Delta \cdot 2c\sqrt{2} - \Delta \cdot 2c \sqrt{n + 1}
\end{align*}
As \(\Delta\) is constant, we conclude that \(E(n) \in O(n)\). 
\end{IEEEproof}

Note that this implies that the reverse operation in which each vertex sends a message to its parent also takes \(O(n)\) energy. Moreover, observe that all the space-filling curves apart from the Z-order mentioned in \Cref{sec:background} satisfy the distance-bound property.

\subsection{Z-Light-First Order}
\begin{figure}
\vspace{-1em}
 \begin{center}
    \resizebox{0.3\linewidth}{!}{
\begin{tikzpicture}[step=0.75cm, x=0.75cm, y=0.75cm]
    \node[proc] at (0.5,2.5){2};
    \node[proc] at (0.5,3.5){0};
    \node[proc] at (1.5,3.5){1};
    \node[proc] at (1.5,2.5){3};
    \node[proc, fill=color2] at (0.5,0.5){10};
    \node[proc, fill=color2] at (0.5, 1.5){8};
    \node[proc, fill=color2] at (1.5, 1.5){9};
    \node[proc] at (1.5, 0.5){11};

    \node[proc, fill=color2] at (2.5,2.5){6};
    \node[proc] at (2.5,3.5){4};
    \node[proc] at (3.5,3.5){5};
    \node[proc, fill=color2] at (3.5,2.5){7};

    \node[proc] at (2.5,0.5){14};
    \node[proc] at (2.5,1.5){12};
    \node[proc] at (3.5,1.5){13};
    \node[proc] at (3.5,0.5){15};

    \begin{scope}[on background layer]
        \draw[help lines, dashed] grid +(4,4);
        \draw[thick] (0.5,1.5) -- ++(1,0) -- ++(-1,-1) -- ++(1,0);
        \draw[thick] (0.5,3.5) -- ++(1,0) -- ++(-1,-1) -- ++(1,0);
        \draw[thick] (2.5,3.5) -- ++(1,0) -- ++(-1,-1) -- ++(1,0);
        \draw[thick] (2.5,1.5) -- ++(1,0) -- ++(-1,-1) -- ++(1,0);
        \draw[thick] (1.5, 2.5) -- (2.5, 3.5);
        \draw[very thick, draw=blue] (3.5, 2.5) -- (0.5, 1.5);
        \draw[thick] (1.5, 0.5) -- (2.5, 1.5);
        \draw[thick, draw=color2] (0.5, 1.5) -- (1.5, 1.5) -- ++(-1, -1);
        \draw[thick, draw=color2] (2.5,2.5) -- (3.5,2.5);
    \end{scope}

\end{tikzpicture}
}
 \end{center}
\vspace{-3em}
  \caption{16 elements stored in Z-order. Given \(i = 6\) and \(j = 10\) the longest diagonal would be the blue one. The x-length of the diagonal would be 3 and the y-length 1. Moreover, we have that \(E_d(6, 10) = 4\).}
  \label{fig:z_order}
\end{figure}
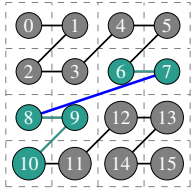

Proving that Z-light-first-order is energy-bound is more involved since, in contrast to the other space-filling curves, it is not distance-bound. This is because the distance between two arbitrary processors is not bounded except by the side lengths of the computational grid. Nevertheless, we will prove the necessary bound on the communication distances by showing that Z-order, in essence, consists of multiple distance-bound curves connected by `diagonals'. The proof relies on showing that each such diagonal is used a bounded number of times.

\begin{theorem}\label{thm:z-order-energy-bound}
  Z-light-first order is energy-bound.
\end{theorem}

If for a space-filling curve for any \(k\) every \(4^k\) consecutive elements are stored in a subgrid of size at most \(2 \cdot 4^\frac{k}{2} \times 2 \cdot 4^\frac{k}{2}\), we call the curve \emph{aligned}. An example of such a curve is the Hilbert curve. We define \(E_b(i,j)\) as the largest energy of sending a message from \(i\)-th to \(j\)-th processor stored on an aligned curve. 

We define \(E_d(i, j)\) as the Manhattan distance of the longest diagonal between \(i\) and \(j\). Refer to \Cref{fig:z_order} for an example. Observe that the Manhattan length of a diagonal is always one larger than the side length of the smallest power-of-two-aligned square subgrid. For simplicity of notation, we define the \emph{length} of a diagonal to be one less than its Manhattan distance.

\begin{lemma}\label{sum}
The energy \(E(i, j)\) of sending a message from \(i\)-th to \(j\)-th processor stored in z-light-first order is at most \(E_b(i,j) + E_d(i, j)\).
\end{lemma}

\begin{IEEEproof}
  In Z-order, every \(4^k\) consecutive and aligned elements are stored in a subgrid of size \(4^\frac{k}{2}\times 4^\frac{k}{2}\) \cite{SpaceFillingBook}. If they are not aligned they would be stored in at most two subgrids of size \(4^\frac{k}{2}\times 4^\frac{k}{2}\) connected by some diagonal and could therefore be far apart. The energy to send a message from \(i\) to \(j\) is then bounded by the energy of sending a message as if those two subgrids were next to each other (all elements stored in \(2 \cdot 4^\frac{k}{2} \times 2 \cdot 4^\frac{k}{2}\)) plus the energy of sending a message across a diagonal. This is bounded by \(E_b(i,j) + E_d(i, j)\).
\end{IEEEproof}

\begin{lemma}\label{third}
An aligned curve is distance bound.
\end{lemma}
\begin{IEEEproof}
The distance between the \(i\)-th and \((i + j)\)-th processor is at most \(8 \sqrt{j}\). Hence, the curve is distance-bound. 
\end{IEEEproof}

Therefore, we conclude by \Cref{third} and \Cref{total message}:
\begin{corollary}\label{energy_b}
The total \(E_b\) cost of each vertex sending a message to all its children is in \(O(n)\).
\end{corollary}


It remains to prove an $O(n)$ bound for the energy of the diagonals \(E_d\). 
Given a diagonal \(D\) of length \(k\), we want to bound how many times it is the longest diagonal when each vertex \(u\) sends a message to its child \(c_i\). We say that \(D\) is the \emph{longest diagonal} in a subtree rooted at some vertex \(v\) if and only if \(D\) is the longest diagonal when \(v\) sends a message to one of its children.
\begin{lemma}\label{diagonal_decrease}
Consider a vertex \(v\) with children \(c_1, ..., c_k\) and let \(D\) be the longest diagonal when \(v\) sends a message to one of its children. Then \(D\) can be the longest diagonal in at most one of the children's subtrees, which has size at most \(\frac{1}{2}\cdot s(u)\).
\end{lemma}
\begin{IEEEproof}
  Note that \(D\) can only be part of one of the subtrees \(c_1, ..., c_{k - 1}\): If \(D\) were in the subtree \(c_k\), then \(v\) would not send a message to any of its children over \(D\). 
  Because only the last subtree may have a size greater than \(\frac{1}{2} \cdot s(u)\), but it is not included,
  \(D\) will be the longest diagonal in a subtree of size at most \(\frac{1}{2}\cdot s(u)\). 
\end{IEEEproof}

\begin{lemma} \label{diagonal}
  Let \(D\) be some diagonal of length \(k = 2^c\), for some \(c \in \mathbb{N^+}\). \(D\) is the longest diagonal at most \(\Delta \cdot \lceil \log_2(4 \cdot k^2)\rceil\) times.
\end{lemma}
\begin{IEEEproof}
  Observe that for every \(4\cdot k^2\) elements, the longest diagonal has a length of at least \(2 \cdot k\). Moreover, diagonals can only have lengths that are powers of 2. Hence, if a diagonal of length \(k\) is the longest diagonal when \(i\) sends a message to \(j\), then \(j - i < 2 \cdot k\).

  Let \(T'\) be the smallest subtree such that for all vertices not in the subtree, \(D\) is never the longest diagonal. It follows that \(T'\) has size at most \(4 \cdot k^2\). Now, observe that every time \(D\) is the longest diagonal in some subtree, the size of the next subtree in which \(D\) can be the longest diagonal is at least halved, by Lemma \ref{diagonal_decrease}. This means that there are at most \(\lceil \log_2(4 \cdot k^2)\rceil\) subtrees in which \(D\) is the longest diagonal. Since each vertex sends at most \(\Delta\) messages, \(D\) is the longest diagonal at most \(\Delta \cdot \lceil \log_2(4 \cdot k^2)\rceil\) times.
\end{IEEEproof}

\begin{lemma}\label{energy_d}
  The total \(E_d\) cost of each vertex sending a message to all its children is in \(O(n)\).
\end{lemma}

\begin{IEEEproof}
  Consider \(n\) processors on a \(\sqrt{n} \times \sqrt n\) grid. For each \(i \in \{0, ..., \lceil\log_2(\sqrt n)\rceil\}\) we have less than \(2 \cdot 4^{\lceil\log_2(\sqrt n)\rceil - i}\) diagonals of length \(2^{i + 1}\), by Lemma \ref{diagonal}. We now bound the diagonal energy:
  \begin{displaymath}
    \begin{split}
      E_d &\leq \sum_{i=0}^{\lceil\log_2\sqrt{n}\rceil} 2 \cdot 4^{\lceil\log_2(\sqrt{n})\rceil - i} \cdot \Delta \cdot 2^{i + 1} \cdot \lceil \log_2(4 \cdot 2^{2i + 2})\rceil\\
      & \leq 16 \cdot \Delta  \sum_{i=0}^{\log_2\sqrt{n} + 1} \frac{n}{2^i}\cdot (4i + 5)\\
      & \leq 8 \cdot \Delta  \sum_{i=1}^{\infty} \frac{n}{2^i}\cdot (4i + 1)
       \leq 8 \cdot \Delta  \cdot n (8 + 1) \in O(n) \\
    \end{split}
  \end{displaymath}
  \vspace{-1em}
\end{IEEEproof}
\Cref{thm:z-order-energy-bound} now follows from Lemmas \ref{sum}, \ref{energy_b} and \ref{energy_d}.

\subsection{Unbounded Degree Trees}\label{sec:unbounded}

So far, we have relied on a restriction to bounded degree trees. When the degree is unbounded, we cannot store all the neighbors of a vertex in one processor because it has $O(1)$ memory. Moreover, energy bounds would deteriorate with the degree even if we allowed that. Instead, we show how to transform the tree of unbounded degree into a tree of bounded degree that can simulate the original tree with \emph{linear energy and using constant space per processor}.

For this to work out, we need to restrict the messaging operations. These operations suffice to implement the two tree algorithms we consider, treefix sum and LCA. We allow so-called \emph{local messaging}, which includes two operations:
\begin{itemize}
  \item \emph{Local broadcast: }Each vertex sends a single message to all its children. Note that the message each child receives has to be the same.
  \item \emph{Local reduce: }Each parent receives a reduction of the messages from its children. The reduction can use any associative function, such as sum or maximum. 
\end{itemize}
We show that for such operations, the total energy cost is \(O(n)\) and the depth is \(O(\log n)\). In the following, light-first order refers to some arbitrary fixed energy-bound light-first order.

We first show how to conceptually transform an unbounded degree tree \(T\) into a tree \(\hat T\) with a bounded degree. \(\hat T\) defines the order in which the messages are sent and propagated. We then show that assuming that \(T\) is in light-first order, \(\hat T\) will also be in light-first order, i.e., we do not have to change the processors in which vertices are stored. Finally, we show how to implement the construction of \(\hat T\), given \(T\).

For a vertex \(v\), we initialize a set \(C(v)\) of \emph{current children} that initially contains all of the children of \(v\) and a set \(A(v)\) of \emph{appended children} that is initially empty.

Let us now assume that we are given a vertex \(v\) with the set of current children \(C(v) = \{c_1, ..., c_d\}\) and the set of appended children \(A(v) = \{a_1, ..., a_{d'}\}\). We define \textsc{Transform(v)}:

\begin{enumerate}[leftmargin=15pt]
  \item Set \(A(c_1)\) to \(\{c_2, ..., c_{\lfloor\frac{d}{2}\rfloor}\}\) and \(A(c_{\lfloor\frac{d}{2}\rfloor + 1})\) to \(\{c_{\lfloor\frac{d}{2}\rfloor + 2}, ..., c_d\}\). Update the set of current children \(C(v)\) to be \(\{c_1, c_{\lfloor\frac{d}{2}\rfloor + 1}\}\).
  \item Set \(A(a_1)\) to \(\{a_2, ..., a_{\lfloor\frac{d'}{2}\rfloor}\}\) and \(A(a_{\lfloor\frac{d'}{2}\rfloor + 1})\) to\(\{a_{\lfloor\frac{d'}{2}\rfloor + 2}, ..., a_{d'}\}\). Update the set of appended children \(A(v)\) to \(\{a_1, a_{\lfloor\frac{d'}{2}\rfloor + 1}\}\).
  \item Transform the vertices \(c_1, c_{\lfloor\frac{d}{2}\rfloor + 1}, a_1\) and \(a_{\lfloor\frac{d'}{2}\rfloor + 1}\).
\end{enumerate}

\begin{figure}
\begin{center}
  \resizebox{.6\linewidth}{!}{
\begin{tikzpicture}[step=.75cm, x=0.75cm, y=0.75cm, node distance=8mm]
    \node (x) {};
    \node[node, font=\small,node distance=15mm] (r) [above of=x]{$v$};
    \node[node distance=10mm,font=\tiny] (dots1)[below of=r]{...};
    \node[nodesmall,node distance=5mm] (a1) [left of=dots1]{$a_1$};
    \draw (a1) -- +(-0.35,-1) -- +(0.35,-1) -- (a1);

    \node[nodesmall] (c2) [left of=a1]{\scalebox{0.6}{$c_{\lfloor\frac{d}{2}\rfloor + 1}$}};
    
    \node[nodesmall, font=\tiny] (c1) [left of=c2]{$c_1$};
    \node[node,inner sep=1,node distance=6mm] (am) [right of=dots1]{\scalebox{0.5}{$a_{\lfloor\frac{d'}{2}\rfloor + 1}$}};
    \node[font=\tiny,node distance=6mm] (dots2)[right of=am] {...};
    \node[nodesmall,font=\tiny,node distance=5mm] (aj) [right of=dots2]{$a_{d'}$};

    \draw (c1) -- +(-0.35,-1) -- +(0.35,-1) -- (c1);
    \draw (c2) -- +(-0.35,-1) -- +(0.35,-1) -- (c2);
    \draw (am) -- +(-0.35,-1) -- +(0.35,-1) -- (am);
    \draw (aj) -- +(-0.35,-1) -- +(0.35,-1) -- (aj);

    \draw (r) -- (c1);
    \draw (r) -- (c2);
    \draw[dashed] (r) -- (a1);
    \draw[dashed] (r) -- (am);
    \draw[dashed] (r) -- (aj);

\end{tikzpicture}
}

\vspace{1em}
\resizebox{.8\linewidth}{!}{
\begin{tikzpicture}[step=.75cm, x=0.75cm, y=0.75cm, node distance=8mm]
  \node[node, font=\small] (r) {$v$};
  \node[nodesmall,node distance=12mm] (a1) [below left of=r]{$a_1$};
  \draw (a1) -- +(-0.35,-1) -- +(0.35,-1) -- (a1);

  \node[nodesmall] (c2) [left of=a1, node distance=8mm]{\scalebox{0.6}{$c_{\lfloor\frac{d}{2}\rfloor + 1}$}};
  
  \node[nodesmall, font=\tiny, node distance=12mm] (c1) [left of=c2]{$c_1$};
  \node[node,node distance=20mm,inner sep=1] (am) [right of=a1]{\scalebox{0.5}{$a_{\lfloor\frac{d'}{2}\rfloor + 1}$}};
  \node[nodesmall,node distance=12mm] (a2) at ($(a1) + (0.8,-1.4)$){$a_2$};
  \node[font=\tiny,node distance=5mm] (dots3)[right of=a2] {...};
  \node[node,node distance=5mm,inner sep=1] (amm) [right of=dots3]{\scalebox{0.5}{$a_{\lfloor\frac{d'}{2}\rfloor}$}};
  \node[node,node distance=5mm,inner sep=1] (amp) at ($(am) + (0.8,-1.4)$){\scalebox{0.5}{$a_{\lfloor\frac{d'}{2}\rfloor + 2}$}};
  \node[font=\tiny,node distance=5mm] (dots2)[right of=amp] {...};
  \node[nodesmall,font=\tiny,node distance=5mm] (aj) [right of=dots2]{$a_{d'}$};
  
  \draw (c1) -- +(-0.35,-1) -- +(0.35,-1) -- (c1);
  \draw (c2) -- +(-0.35,-1) -- +(0.35,-1) -- (c2);
  \draw (am) -- +(-0.35,-1) -- +(0.35,-1) -- (am);
  \draw (aj) -- +(-0.35,-1) -- +(0.35,-1) -- (aj);
  \draw (a2) -- +(-0.35,-1) -- +(0.35,-1) -- (a2);
  \draw (amm) -- +(-0.35,-1) -- +(0.35,-1) -- (amm);
  \draw (amp) -- +(-0.35,-1) -- +(0.35,-1) -- (amp);

  \draw (r) -- (c1);
  \draw (r) -- (c2);
  \draw[dashed] (r) -- (a1);
  \draw[dashed] (r) -- (am);
  \draw[dashed] (am) -- (aj);
  \draw[dashed] (a1) -- (a2);
  \draw[dashed] (a1) -- (amm);
  \draw[dashed] (am) -- (amp);
\end{tikzpicture}
}
\end{center}
  \vspace{-3em}
  \caption{Example of a virtual tree rooted at a vertex \(v\) before and after the second step of \textsc{Transform}. The vertex $v$ has degree $4$ after the transform. Solid lines connect to the current children, whereas dashed lines connect to the appended children.}
  \label{fig:virtual_tree}
\end{figure}
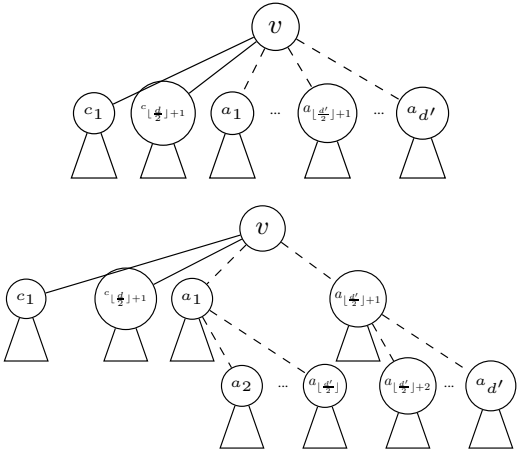

See \Cref{fig:virtual_tree} for an example of the \textsc{Transform}. We define a \emph{virtual} tree as a tree where the children of a vertex \(v\) are a union of \(C(v)\) and \(A(v)\). Now let \(\hat T\) be a virtual tree which at first is the same as \(T\). We then use the algorithm defined above to conceptually transform \(\hat T\) starting from the root. The final result is a binary virtual tree \(\hat T\) which defines the messaging order.

We now show that after the transformation, the children of every vertex are still sorted by the size of their subtree. This means that the physical placement of the vertices does not need to change:
\begin{lemma}\label{temp tree} Let \(\hat T\) be a tree resulting from transformation of the tree \(T\). If \(T\) is in light-first order and the vertices in \(\hat T\) are stored in the same order as in \(T\), then \(\hat T\) is also in light-first order.
\end{lemma}
\begin{IEEEproof}
We assume that the vertices in \(\hat T\) are stored in the same processors as in \(T\). We prove the statement by induction. \(\hat T\) is in light-first order at the beginning since it is the same as \(T\). Let \(T\) be the state of \(\hat T\) before some step of the transformation and \(T'\) the state of \(\hat T\) after that step. We now show that assuming a tree \(T\) is in light-first order, the tree \(T'\) is also in light-first order. We consider some vertex \(v\) and first look at step 1. 
Let \(C(v) = \{c_1, ..., c_d\}\) and \(A(v) = \{a_1, ..., a_{d'}\}\). We execute step 1 of \textsc{Transform(v)}. The correct positions of the vertex \(v\) and the vertices outside of the subtree rooted at \(v\) stay the same since the size of the subtree rooted at \(v\) does not change. Notice that the children of \(v\) stay sorted by size, since \(|A(c_1)| \leq |A(c_{\lfloor\frac{d}{2}\rfloor + 1})|\) and \(s(c_i) \leq s(c_j)\) for all \(i \leq j\) before the operation. It therefore follows that \(s(c_1) \leq s(c_{\lfloor\frac{d}{2}\rfloor + 1})\) after the operation. 
Note that step 2 can be analyzed analogously. In particular, if $A(v)$ is not empty, then \(s(c_{\lfloor\frac{d}{2}\rfloor + 1}) \leq s(a_1) \leq s(a_{\lfloor\frac{d'}{2}\rfloor + 1})\) after the operation. Since the vertices remain sorted by their subtree size, the positions in which the subtrees need to be stored stay the same. As the algorithm only executes those steps, the virtual tree \(\hat T\) is always in light-first order.
\end{IEEEproof}

We now assume that we are given the tree \(\hat T\) as input and it is in light-first order. Note that \(\hat T\) has degree bounded by 4. Let \(v \in B\) with \(C(v) = \{c_1, c_2\}\) and \(A(v) = \{a_1, a_2\}\). For each vertex \(v\) we define the local broadcast:
\begin{enumerate}
  \item Send the message to its children \(c_1\) and \(c_2\).
  \item Unless $v$ is the root, wait until $v$ receives a message from the parent and then propagate it to \(a_1\) and \(a_2\).
\end{enumerate}

The local reduce can be defined in a similar way using reduction. 

\begin{lemma} \label{temp messaging}
  Local messaging in a virtual tree \(\hat T\) stored in energy-bound light-first order takes \(O(n)\) energy and \(O(\log n)\) depth.
\end{lemma}

\begin{IEEEproof}
  Since the tree has constant degree and is in energy-bound light-first order it follows from Theorem \ref{total message} that the local messaging is done with \(O(n)\) energy. The depth follows from the fact that if we look at a single vertex we are doubling the number of vertices that distribute its message every round. After \(O(\log n)\) rounds all the messages will therefore be sent.
\end{IEEEproof}

Note that we have shown this assuming that the virtual tree \(\hat T\) is given as input. We now show how to construct \(\hat T\) given some unbounded degree tree \(T\) as input. Note that every processor has $O(1)$ memory, so a vertex cannot store a reference to all its children. 
Next, we show how each vertex can compute its parent and children in the virtual tree \(\hat T\). Each vertex has at most two current children, so it can pass their references to its parent. Next, we focus on the appended children.

\begin{figure}
  \begin{tikzpicture}[scale=0.78, every node/.style={transform shape},step=0.75cm, x=0.75cm, y=0.75cm, node distance=17mm]

    \node[node] (cj) {$c_j$};
    
    \node[node,inner sep=1] (cjp) [below left of=cj] {\scalebox{0.8}{$c_{j+1}$}};
    \node[node] (ck) [below right of=cj] {$c_k$};
    \node[node,inner sep=1,node distance=14mm] (ckm) [below of=cjp]{\scalebox{0.8}{$c_{k-1}$}};
    \node[node,inner sep=1,node distance=14mm] (clm) [below of=ck]{\scalebox{0.8}{$c_{l-1}$}};
    \node[node, node distance=25mm] (cl) [right of=cj] {$c_l$};
    \draw (cj) -- (cjp);
    \draw (cj) -- (ck);
    
    \draw[->,thick,draw=blue] (ck) -- (cl);
    \draw[<->,thick,draw=blue] (clm) -- (cl);
    \draw[<->,thick,draw=blue] (ckm) -- (ck);
    \draw[<->,thick,draw=blue] (cjp) to [out=70,in=190] (cj);
    \draw[->,thick,draw=blue] (cjp) -- (ck);
    \begin{scope}[on background layer]
      \draw[fill opacity=0.6] (cjp) -- +(-1.5,-2.5) -- +(1.5,-2.5) -- (cjp);
      \draw[fill opacity=0.6] (ck) -- +(-1.5,-2.5) -- +(1.5,-2.5) -- (ck);
    \end{scope}
      
    \end{tikzpicture}
    \begin{tikzpicture}[scale=0.78, every node/.style={transform shape},step=0.75cm, x=0.75cm, y=0.75cm, node distance=17mm]
  
      \node[node] (cj) {$c_j$};
      
      \node[node,inner sep=1] (cjp) [below left of=cj] {\scalebox{0.8}{$c_{j+1}$}};
      \node[node] (ck) [below right of=cj] {$c_k$};
      \node[node,inner sep=1,node distance=14mm] (ckm) [below of=cjp]{\scalebox{0.8}{$c_{k-1}$}};
      \node[node,inner sep=1,node distance=14mm] (clm) [below of=ck]{\scalebox{0.8}{$c_{l-1}$}};
      \node[node, node distance=25mm] (cl) [right of=cj] {$c_l$};
      \draw (cj) -- (cjp);
      \draw (cj) -- (ck);
      
      \draw[-stealth,thick,draw=blue,draw opacity=0.6] (ck) -- (cl);
      \draw[<->,thick,draw=blue,draw opacity=0.6] (clm) -- (cl);
      \draw[<->,thick,draw=blue,draw opacity=0.6] (ckm) -- (ck);
      \draw[<->,thick,draw=blue,draw opacity=0.6] (cjp) to [out=70,in=190] (cj);
      \draw[<->,very thick,draw=blue] (ck) to [out=100,in=340] (cj);
      \draw[->,thick,draw=blue,draw opacity=0.6] (cjp) -- (ck);
      \draw[->,very thick,draw=blue] (cj) -- (cl);
      \begin{scope}[on background layer]
        \draw[fill opacity=0.6] (cjp) -- +(-1.5,-2.5) -- +(1.5,-2.5) -- (cjp);
        \draw[fill opacity=0.6] (ck) -- +(-1.5,-2.5) -- +(1.5,-2.5) -- (ck);
      \end{scope}
        
      \end{tikzpicture}
    \vspace{-2em}
  \caption{Example of a procedure for passing the references. The black edges represent the tree \(\hat T\). The directed blue edges represent each vertex's references before and after the operation.}
  \label{fig:references}
\end{figure}
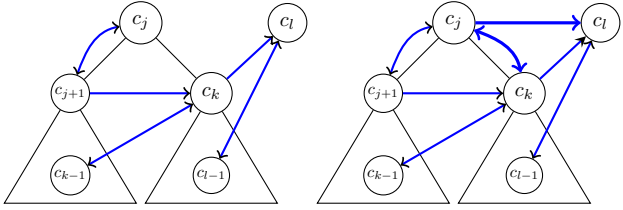

Let \(v \in U\) with \(C(v) = \{c_1, ..., c_d\}\). We assume that each \(c_j \in C(v)\) knows its index $j$ and the degree $d$ of its parent. Moreover, \(c_j\) has a reference to its left sibling \(c_{j - 1}\), its right sibling $c_{j + 1}$, and its parent \(v\). 
Assume that \(A(c_j) = \{c_{j + 1}, c_k\}\) for some $k$ and that $c_j$'s parent is \(c_p\). Observe that $c_k$ is the right sibling (in the list of children of $v$) of the rightmost descendant of $c_j$. 
The procedure works bottom-up in the subtree rooted at the appended children and maintains this right sibling of the rightmost descendant of the current subtree. 
Assume that \(c_{j + 1}\) is either a leaf or has finished local messaging with its children. Then, \(c_{j + 1}\) sends \(c_j\) the reference to \(c_k\). If \(c_{j + 1}\) is a leaf then \(c_k = c_{j + 2}\) hence it already has that reference. Otherwise, \(c_{k - 1}\) is in the subtree rooted at \(c_{j + 1}\). Therefore, \(c_{j + 1}\) would have received that reference. Next, \(c_j\) sends a message to \(c_k\) which responds with the reference to \(c_l\), where \(c_l\) is the only vertex that is not in the subtree rooted at \(c_k\) where \(c_{l - 1}\) is in that subtree. See Figure~\ref{fig:references} for an illustration. 
To compute the parent $c_p$, observe that if $c_j$ is a left child of its parent, then $c_p=c_{j-1}$. Otherwise, $c_j$ waits until it gets the reference to the parent from its sibling. 
Note that finding the references to the children and parents is enough to build the virtual \(\hat T\) because, as proven in Lemma~\ref{temp tree}, we do not have to change the position of the vertices.

\begin{theorem} \label{unbounded messaging}
  Local messaging in a tree \(T\) stored in energy-bound light-first order takes \(O(n)\) energy and \(O(\log n)\) depth.
\end{theorem}

\begin{IEEEproof}
We construct \(\hat T\) as shown above. This requires \(O(\log n)\) depth and \(O(1)\) memory for each vertex to compute its children and its parent. It then takes \(O(n)\) energy and \(O(\log n)\) depth to pass the references, which follows from Lemma \ref{temp messaging}. Once we have constructed \(\hat T\), every local messaging operation takes \(O(n)\) energy and \(O(\log n)\) depth.
\end{IEEEproof}

\section{Creating the Layout}

We have shown how energy-bound light-first order is a storage format that leads to linear-energy messaging within the tree. 
It remains to show how a light-first order tree layout can be efficiently computed. 
Our goal is to obtain \(O(n^{\frac{3}{2}})\) energy, which matches the permutation lower bound, and $O(\log n)$ depth. The main challenge lies in reducing the depth: 
Simulating a work-optimal PRAM algorithm would give \(\Theta(n^{\frac{3}{2}})\) energy and \(\Theta(\log^4 n)\) depth~\cite{SpatialModel}.

\begin{theorem}\label{thm:light-first-order}
Light-first order takes \(O(n^{\frac{3}{2}})\) energy and \(O(\log n)\) depth to compute, with high probability.
\end{theorem}

The approach is as follows:
\begin{enumerate}[leftmargin=14pt]
    \item Compute the size of each subtree via an Euler Tour~\cite{84TarjanEulerTour}:
    \begin{enumerate}
        \item Drop all but the first and last occurrence of a vertex using a parallel prefix sum and compact the result. 
        \item The size of a vertex $v$'s subtree is half the difference between the first and last index of $v$ in the tour.
    \end{enumerate}
    \item Create an Euler Tour of the tree, where the children are visited in increasing order of their subtree size.
    \item Drop all but the first occurrence of each vertex using a parallel prefix sum.
    \item Permute the vertices according to a space-filling curve.
\end{enumerate}

Next, we show to compute an Euler Tour using list ranking.  
List ranking is the problem of determining the index of each element in a linked list. By duplicating every edge in a rooted tree and ranking the resulting list, it computes an Euler Tour.

%
We adapt a contraction-based algorithm~\cite{DBLP:journals/ipl/AndersonM90}. The idea is to repeatedly contract a large independent set of edges. 

Initially, each processor stores the index of one vertex and one pointer to the processor that stores the next vertex in the list. 
To compute the list ranking, for a given list, first select a subset $S$ of non-adjacent vertices using random-mate~\cite{DBLP:journals/ipl/AndersonM90, pem_lca, ReifSynthesis}. Then, do one step of pointer jumping over the vertices from $S$. When doing the pointer jumping step, store the information of the current iteration number. Increase the temporary vertex rank of any vertex that has its pointer adjusted by the temporary rank of the vertex of which it copies the pointer. 
Once $\Theta(\log n)$ vertices remain, solve the list ranking problem sequentially. Finally, revert each step of the algorithm, where in each step the vertices that were in $S$ during that iteration, compute their local rank with respect to the vertex that copied their pointer.
\begin{theorem}
We can compute the list ranking of a list with $n$ vertices with ${O}(n^\frac{3}{2})$ energy and ${O}(\log n)$ depth, with high probability.
\end{theorem}
\begin{IEEEproof}[Proof Sketch]
As long as $\Omega(\log n)$ vertices remain, the number of edges decreases by a constant factor in each iteration by a Chernoff bound~\cite{chernoff}. The energy of an iteration where $n'$ vertices remain is $O(n'\sqrt{n})$ and its depth is $O(1)$. The base case takes ${O}(\sqrt{n} \log n)$ energy and ${O}(\log n)$ depth.
\end{IEEEproof}

As the bottleneck of computing the Euler Tour is in the list ranking, we conclude:
\begin{corollary}\label{cor:euler-tour}
    Computing an Euler tour on a tree with $n$ vertices takes $\mathcal{O}(n^\frac{3}{2})$ energy and $\mathcal{O}(\log n)$ depth with high probability.
\end{corollary}

Since all the operations involved in creating the layout can be reduced to the computation of Euler tours, sorting, and parallel prefix sums, \Cref{thm:light-first-order} follows. We proceed to show two tree algorithms using our energy-efficient layout.


\section{Treefix Sum}\label{sec:treefix}

We showcase the application of our tree layouts to a classic algorithmic problem. This problem generalizes prefix sums and is related to the parallel evaluation of arithmetic expressions~\cite{MiRe85}. Given a rooted tree $T$ with a value in each node, the goal of a \emph{treefix sum} is to compute for each vertex $v$ the sum of the values in the subtree rooted at $v$. Any associative operator may be used instead of a sum. This problem has applications in minimum cut computations~\cite{KargerMinCut, LGianCut, DBLP:conf/spaa/AndersonB21}.

We present a Las Vegas algorithm which solves the treefix sum problem with \(O(n \log n)\) energy and \(O(\log ^2 n)\) depth. The algorithm has \(O(\log n)\) depth for trees of bounded degree. The input is a tree \(T\) stored in energy-bound light-first order. Each vertex \(u\) initially holds a value \emph{val}\((u)\). After the algorithm finishes, each vertex \(v\) holds the sum  \emph{sum}\((v)\) of the values of its descendants, including its own value \emph{val}\((v)\). 

The algorithm has two phases, \emph{tree contraction} and a \emph{uncontraction}. In the tree contraction phase, we use modified versions of the \emph{rake} and \emph{compress} operations~\cite{MiRe85} and maintain partial sums over merged subtrees. After contracting the whole tree, we undo the contractions to compute the final results using those partial sums. Using this framework ensures that the algorithm has low depth for any tree. 
The challenges are twofold. First, we need to use local messaging to ensure energy-efficiency. Second, we need to maintain the state using only constant memory per processor.




\subsection{Tree Contraction}
\begin{figure}
\vspace{-1em}
\begin{center}
    \begin{tikzpicture}[scale=0.73, every node/.style={transform shape},step=0.75cm, x=0.75cm, y=0.75cm, node distance=10mm]

    \node[node,fill=color1] (u) {$u$};
  
    \node[child2] (v) [below of=u] {$v$};
    \node[child1] (w) [below of=v] {$w$};
  
    \draw (u) -- (v);
    \draw (v) -- (w);
  
    \draw[-stealth] ($(v) + (0.5,0)$) -- +(0.5,0);
    
  \end{tikzpicture}
  \begin{tikzpicture}[scale=0.73, every node/.style={transform shape},step=0.75cm, x=0.75cm, y=0.75cm, node distance=10mm]

    \fill[fill=color1light] (0,0) rectangle (1,1);
  
    \fill[fill=color1light] (1,0) rectangle (2,1);
    \fill[fill=color1light] (2,1) rectangle (1,2);
    \fill[fill=color2light,] (0,1) rectangle (1,2);
    \fill[fill=color2light,] (0,2) rectangle (1,4);
    \fill[fill=color2light,] (1,4) rectangle (2,3);
  
    \fill[fill=color3light] (1,3) rectangle (2,2);
    \fill[fill=color3light] (2,4) rectangle (4,0);

    \draw[help lines, dashed] grid +(4,4);
  
    \draw[thick, color = gray] (0.5,0.5) -- ++(1, 0) -- ++(0,1) -- ++(-1,0) -- ++(0,1);
    \draw[thick, color = gray] (0.5,2.5) -- ++(0,1) -- ++(1,0) -- ++(0,-1) -- ++(1,0);
    \draw[thick, color = gray] (2.5,2.5) -- ++(0,1) -- ++(1,0) -- ++(0,-1) -- ++(0,-1);
    \draw[thick, color = gray] (3.5,1.5) -- ++(-1,0) -- ++(0,-1) -- ++(1,0);
    
    \node[node,fill=color1] (r) at (0.5,0.5){$u$};
    \node[child1,font=\normalsize,inner sep=3] (c1) at (0.5,1.5){$v$};
    \node[child2,font=\small,inner sep=2.25] (c2) at (1.5,2.5){$w$};
    \draw[->,thick] (r) -- (c1);
    \draw[->,thick] (c1) -- (c2);
  
    \draw[-stealth] (4.25,2) -- +(0.5,0);

  \end{tikzpicture}
  \begin{tikzpicture}[scale=0.73, every node/.style={transform shape},step=0.75cm, x=0.75cm, y=0.75cm, node distance=10mm]

    \fill[fill=color1light] (0,0) rectangle (1,1);
  
    \fill[fill=color1light] (1,0) rectangle (2,1);
    \fill[fill=color1light] (2,1) rectangle (1,2);
    \fill[fill=color1light,] (0,1) rectangle (1,2);
    \fill[fill=color1light,] (0,2) rectangle (1,4);
    \fill[fill=color1light,] (1,4) rectangle (2,3);
  
    \fill[fill=color3light] (1,3) rectangle (2,2);
    \fill[fill=color3light] (2,4) rectangle (4,0);

    \draw[help lines, dashed] grid +(4,4);
  
    \draw[thick, color = gray] (0.5,0.5) -- ++(1, 0) -- ++(0,1) -- ++(-1,0) -- ++(0,1);
    \draw[thick, color = gray] (0.5,2.5) -- ++(0,1) -- ++(1,0) -- ++(0,-1) -- ++(1,0);
    \draw[thick, color = gray] (2.5,2.5) -- ++(0,1) -- ++(1,0) -- ++(0,-1) -- ++(0,-1);
    \draw[thick, color = gray] (3.5,1.5) -- ++(-1,0) -- ++(0,-1) -- ++(1,0);
    
    \node[node,fill=color1] (r) at (0.5,0.5){$u$};
    \node[child2,font=\small,inner sep=2.25] (c2) at (1.5,2.5){$w$};
    \draw[->,thick] (r) -- (c2);
    \draw[-stealth] (4.25,2) -- +(0.5,0);
  
  \end{tikzpicture}
  \begin{tikzpicture}[scale=0.73, every node/.style={transform shape},step=0.75cm, x=0.75cm, y=0.75cm, node distance=10mm]

    \fill[fill=color1light] (0,0) rectangle (1,1);
  
    \fill[fill=color1light] (1,0) rectangle (2,1);
    \fill[fill=color1light] (2,1) rectangle (1,2);
    \fill[fill=color1light,] (0,1) rectangle (1,2);
    \fill[fill=color1light,] (0,2) rectangle (1,4);
    \fill[fill=color1light,] (1,4) rectangle (2,3);
  
    \fill[fill=color1light] (1,3) rectangle (2,2);
    \fill[fill=color1light] (2,4) rectangle (4,0);

    \draw[help lines, dashed] grid +(4,4);
  
    \draw[thick, color = gray] (0.5,0.5) -- ++(1, 0) -- ++(0,1) -- ++(-1,0) -- ++(0,1);
    \draw[thick, color = gray] (0.5,2.5) -- ++(0,1) -- ++(1,0) -- ++(0,-1) -- ++(1,0);
    \draw[thick, color = gray] (2.5,2.5) -- ++(0,1) -- ++(1,0) -- ++(0,-1) -- ++(0,-1);
    \draw[thick, color = gray] (3.5,1.5) -- ++(-1,0) -- ++(0,-1) -- ++(1,0);

    \node[node,fill=color1] (r) at (0.5,0.5){$u$};
  
  \end{tikzpicture}
    \vspace{-3em}
\end{center}
  \caption{Example illustrating the compression of supervertices. Every supervertex corresponds to several vertices as indicated by the contiguous space taken up in the grid drawing. We first compress \(u\) with \(v\) and then \(u\) with \(w\).}
  \label{fig:Compress}
\end{figure}
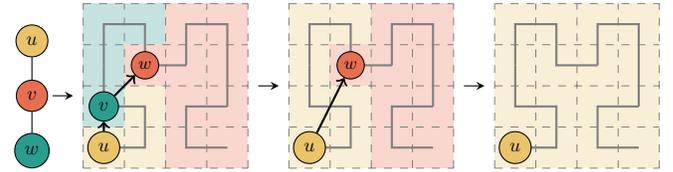

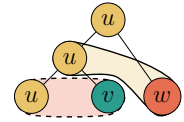
\begin{figure}
\begin{center}
    \resizebox{0.27\linewidth}{!}{
      \begin{tikzpicture}[step=0.75cm, x=0.75cm, y=0.75cm, node distance=11mm]
        \node[node,font=\Large,fill=color1] (u1) {$u$};
        \node[node,font=\Large,fill=color1] (u2) [below left of=u1]{$u$};
        \node[node,font=\Large,fill=color1] (u3) [below left of=u2]{$u$};
        \node[node,font=\Large,fill=color2] (v) [below right of=u2]{$v$};
        \node[node,font=\Large,fill=color3] (w) [right of=v]{$w$};
    
        \draw (u1) -- (u2);
        \draw (u2) -- (u3);
        \draw (u2) -- (v);
        \draw (u1) -- (w);
    
        \begin{scope}[on background layer]
          \draw [thick, dashed,fill=color3light] plot [smooth cycle] coordinates {($(u3) + (0,+0.45)$) ($(u3) + (-0.45,0)$) ($(u3) + (0,-0.5)$) ($(v) + (0,-0.5)$) ($(v) + (0.45,0)$) ($(v) + (0,0.45)$)};
    
          \draw [thick,fill=color1light] plot [smooth cycle] coordinates {($(u2) + (0,+0.45)$) ($(u2) + (-0.45,0)$) ($(u2) + (-0.1,-0.45)$) ($(u2) + (1.25,-0.5)$)($(w) + (-0.15,-0.5)$) ($(w) + (0.45,-0.2)$) ($(w) + (0.25,0.45)$) ($(w) + (-0.75, 1)$)};
    
        \end{scope}
      \end{tikzpicture}
    }
\end{center}
\vspace{-0.5em}
  \caption{Example illustrating a contraction tree. The tree is a result of the contractions on the supervertices shown in Figure \ref{fig:Compress}. The solid border represents \emph{last\_contracted}$(u)$, whereas the dashed one \emph{saved\_state(w)}
  .}
  \label{fig:Contraction}
\end{figure}

We adapt the classic rake and compress~\cite{MiRe85} to our spatial setting. 
A subset of vertices that induces a single connected component in the tree is a \emph{supervertex}. A supervertex $u$ is represented by and identified with its vertex closest to the root, called its \emph{representative} $R(u)$. This is the first vertex in light-first order of the subgraph induced by the supervertex. Throughout, we will maintain the invariant that every supervertex $u$ stores the \emph{partial sum} $P_u$ of its values in its representative $R(u)$. During the contraction process, we consider the tree of supervertices, which is the tree we get when we merge all vertices in the supervertices. We can contract two supervertices and merge their sets of vertices when they are adjacent in the tree of supervertices. Doing so merges their sets of vertices. 

When merging two supervertices, we need to maintain their children before and after the merge. 
This allows us to undo the sequence of contractions later. We want to store the contraction tree (see \Cref{fig:Contraction}), which shows the order in which the supervertices were merged. 
Simply storing a list of operations in the vertex representing the supervertex would lead to unbounded storage. Instead, we distribute this list among the vertices in the supervertex. Each supervertex $u$ stores the representatives of the supervertices that were \emph{last} contracted to create $u$ in its representative $R(u)$ as \emph{last\_contracted}$(u)$. If \emph{last\_contracted}$(u)$ is non-empty, then the previous value of \emph{last\_contracted} before the contraction is stored in \emph{saved\_state}$(\text{\emph{last\_contracted}}(u)[0])$. As long as \emph{last\_contracted}$(u)$ has constant size, this ensures the storage remains $O(1)$ throughout. Similarly, we maintain which type of contraction operation was used to create a supervertex (\textsc{Rake} or \textsc{Compress}).





\subsubsection{Compress}\label{sec:compress}
A \textsc{Compress} operation on the supervertices \(u\) and \(v\) can be performed when \(v\) is \emph{the only} child of \(u\) and $u$ has a single child as well. Such a compress operation contracts \(u\) and \(v\). We increment the partial sum $P_u$ of the supervertex $u$ by the partial sum $P_v$ of the supervertex $v$. 
After that, \(u\) inherits the children of \(v\). 
Finally, we set \(v\) to be \emph{inactive}, meaning it will wait for the uncontraction to be reactivated. This is illustrated in Figure \(\ref{fig:Compress}\).


\subsubsection{Rake}\label{sec:rake}
\begin{figure}
  \begin{tikzpicture}[scale=0.82, every node/.style={transform shape},step=0.75cm, x=0.75cm, y=0.75cm]

    \node[node,fill=color1] (u) {$u$};
  
    \node[child2] (c2) [below of=u] {$v_2$};
    \node[child1] (c1) [ left of=c2] {$v_1$};
  \node[child3] (c3) [ right of=c2] {$w$};
  \node[node,fill=lightgray] (w) [below of=c3]{$x$};
  
  \draw (u) -- (c1);
  \draw (u) -- (c2);
  \draw (u) -- (c3);
  \draw (c3) -- (w);
  
  \draw[-stealth] ($(c3) + (0.75,0)$) -- +(0.5,0);
    \end{tikzpicture}
    \begin{tikzpicture}[scale=0.82, every node/.style={transform shape},step=0.75cm, x=0.75cm, y=0.75cm, node distance=10mm]

        \fill[fill=color1light] (0,0) rectangle (1,1);
      
        \fill[fill=color1light] (1,0) rectangle (2,1);
        \fill[fill=color2light,] (0,1) rectangle (2,2);
        \fill[fill=color2light,] (0,2) rectangle (1,3);
        \fill[fill=color4light,] (0,4) rectangle (2,3);
        \fill[fill=color4light,] (1,3) rectangle (3,2);
        \fill[fill=verylightgray] (2,4) rectangle (3,3);
        \fill[fill=color3light,] (3,4) rectangle (4,0);
        \fill[fill=color3light,] (2,2) rectangle (3,0);

        \draw[help lines, dashed] grid +(4,4);
      
        \draw[thick, color = gray] (0.5,0.5) -- ++(1, 0) -- ++(0,1) -- ++(-1,0) -- ++(0,1);
        \draw[thick, color = gray] (0.5,2.5) -- ++(0,1) -- ++(1,0) -- ++(0,-1) -- ++(1,0);
        \draw[thick, color = gray] (2.5,2.5) -- ++(0,1) -- ++(1,0) -- ++(0,-1) -- ++(0,-1);
        \draw[thick, color = gray] (3.5,1.5) -- ++(-1,0) -- ++(0,-1) -- ++(1,0);
        
        \node[node,fill=color1] (r) at (0.5,0.5){$u$};
        \node[child1,font=\normalsize,inner sep=3] (c1) at (1 .5,1.5){$v_1$};
        \node[child2,font=\small,inner sep=2.25] (c2) at (3.5,3.5){$v_2$};
        \node[child3,font=\small,inner sep=2.25] (c3) at (0.5,3.5){$w$};
        \node[node,font=\small,fill=lightgray,inner sep=2.25] (w) at (2.5,3.5){$x$};
        \draw[->,thick] (r) -- (c1);
        \draw[->,thick] (r) to [out=0,in=270] (c2);
        \draw[->,thick] (r) -- (c3);
        \draw[->,thick] (c3) -- (w);
        \draw[-stealth] (4.25,2) -- +(0.5,0);

      \end{tikzpicture}
      \begin{tikzpicture}[scale=0.82, every node/.style={transform shape},step=0.75cm, x=0.75cm, y=0.75cm, node distance=10mm]

        \fill[fill=color1light] (0,0) rectangle (1,1);
      
        \fill[fill=color1light] (1,0) rectangle (2,1);
        \fill[fill=color1light,] (0,1) rectangle (2,2);
        \fill[fill=color1light,] (0,2) rectangle (1,3);
        \fill[fill=color4light,] (0,4) rectangle (2,3);
        \fill[fill=color4light,] (1,3) rectangle (3,2);
        \fill[fill=verylightgray] (2,4) rectangle (3,3);
        \fill[fill=color1light,] (3,4) rectangle (4,0);
        \fill[fill=color1light,] (2,2) rectangle (3,0);

        \draw[help lines, dashed] grid +(4,4);
      
        \draw[thick, color = gray] (0.5,0.5) -- ++(1, 0) -- ++(0,1) -- ++(-1,0) -- ++(0,1);
        \draw[thick, color = gray] (0.5,2.5) -- ++(0,1) -- ++(1,0) -- ++(0,-1) -- ++(1,0);
        \draw[thick, color = gray] (2.5,2.5) -- ++(0,1) -- ++(1,0) -- ++(0,-1) -- ++(0,-1);
        \draw[thick, color = gray] (3.5,1.5) -- ++(-1,0) -- ++(0,-1) -- ++(1,0);
        
        \node[node,fill=color1] (r) at (0.5,0.5){$u$};
        \node[child3,font=\small,inner sep=2.25] (c3) at (0.5,3.5){$w$};
        \node[node,font=\small,fill=lightgray,inner sep=2.25] (w) at (2.5,3.5){$x$};
        \draw[->,thick] (r) -- (c3);
        \draw[->,thick] (c3) -- (w);

      \end{tikzpicture}
    \vspace{-2.6em}
  \caption{
  The example illustrates the rake of the supervertex \(u\) and its children \(v_1\) and \(v_2\).  Note that \(w\) is not part of the rake since it is not a leaf.}
  \vspace{-1em}
  \label{fig:rake}
\end{figure}
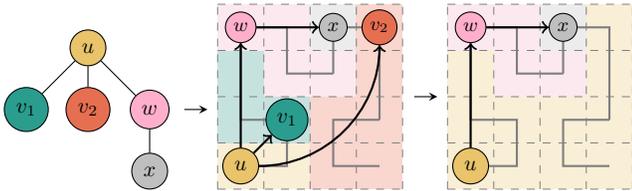

A \textsc{Rake} operation (see Figure~\ref{fig:rake}) on \(u\) and a subset of its children \(v_1, ..., v_i\) can be performed when all these children are leaves and \(u\) has at most one other child $w$. Such a rake operation contracts \(u\) and \(v_1, ..., v_i\) .

Observe that when we perform such a \textsc{Rake}, the supervertex $u$ consists of just a single vertex that is the direct parent of the vertices that represent the supervertices \(v_1, ..., v_i\). Hence, we can perform local messaging to efficiently update the state even when the degree is unbounded. 
During the rake, we increase the partial sum $P_u$ of the supervertex $u$ by the value of a local reduce with an addition operator, where each child \(v\) except $w$ sends its partial sum $P_v$. A child \(w\) that is not part of the rake, if it exists, sends the value $0$. The supervertices \(v_1, ..., v_i\) are set to \emph{inactive}. 

We assumed that \emph{last\_contracted} has constant size. Since \(\deg(u)\) might be unbounded, we cannot store a list of all the vertices that were part of the rake. Instead, we store the representative of the vertex $w$ that \emph{was not} part of the rake.  This is sufficient to undo the \textsc{Rake}, by doing a local broadcast that omits $R(w)$.


\subsubsection{Compact}

The \textsc{Compact} subroutine iteratively applies \textsc{Compress} to reduce the length of the paths in the tree and then \textsc{Rake} to contract the leaves. 
A supervertex $v$ is \emph{viable} if and only if its parent is \emph{non-branching} (i.e., it is the only child) and $v$ has exactly one child. Only the \emph{viable} supervertices can be compressed with their parents.
\textsc{Compact} works as follows:
\begin{enumerate}
    \item Each supervertex sends a message to its child supervertices indicating if it is \emph{branching}. 
    \item Find a set $R$ of independent viable supervertices. This set can be found using local messaging with a randomized approach called random-mate~\cite{DBLP:journals/ipl/AndersonM90, pem_lca, ReifSynthesis}: Independently, each vertex chooses heads or tails with probability $p=\frac{1}{2}$. Include in $R$ every viable vertex that chose heads and whose predecessor chose tails.
    \item \textsc{Compress} every vertex in $R$ with its parent.
    \item Repeat step (1).
    \item \textsc{Rake} the supervertices where it is possible.
\end{enumerate}

\begin{lemma}\label{bough analysis constant}
\textsc{Compact} takes \(O(n)\) energy and its depth is \(O(1)\) or \(O(\log n)\) for bounded and unbounded degree trees, respectively.
\end{lemma}
\begin{IEEEproof}
  ({i}) We first note that by the triangle inequality, the energy of sending a message from \(u\) to \(v\) and then from \(v\) to \(w\) is at least the energy of sending a message from a supervertex \(u\) to a supervertex \(w\) (which was created by contracting \(v\) and \(w\)). (ii) In every round, each supervertex sends a constant number of messages to its children and parents. 
  (iii) By \Cref{unbounded messaging}, the total energy of every vertex sending a message to its parent and children is \(O(n)\). From (i), (ii), and (iii) it follows that 
  the total energy spent per round of \textsc{Compact} is in \(O(n)\). The depth follows from the local messaging bounds in \Cref{total message} and \Cref{unbounded messaging}. 
\end{IEEEproof}

\subsection{Uncontraction}
We can use the partial sums and the contraction tree to undo the contractions and compute the result. We undo the contractions using local messaging based on \emph{last\_contracted} and \emph{saved\_state}. 

We maintain a value $A_u$ for every vertex $u$ that represents the sum of the values of descendants of the current supervertex represented by $u$ that are not in $u$. This value needs to be added to the partial sum of the supervertex $u$ to get the sum at $u$. The invariant is that for every supervertex $u$, its result is \emph{sum}$(u)=P_u + A_u$. Initially, $A_u=0$. This invariant can be maintained as follows. When undoing a \textsc{Compress} of supervertex $u$ with child supervertex $v$, increment $A_u$ by $P_v$ and decrement $P_u$ by $P_v$. For the supervertex $v$, set $A_v=A_u$. 
When undoing a \textsc{Rake} of supervertex $u$ with children $v_1, \dotsc v_i$, set $P_u=\text{\emph{val(u)}}$ and $A_u=P_u-\text{\emph{val}}(u)$.

\subsection{A Local Messaging Treefix Sum Algorithm}\label{bough algorithm}

The local messaging treefix sum algorithm works as follows: As long as there exist two or more supervertices, apply \textsc{Compact} to the supervertices. Once there is one supervertex left, perform uncontraction to get the results. 
Synchronization between the rounds would be a bottleneck (in the bounded degree case) since it requires \(O(\log n)\) depth. Note that in each round of \textsc{Compact}, a supervertex only needs to communicate with its parent and children. Therefore, we can safely avoid global synchronization by having each vertex execute the steps as soon as possible. 

\begin{lemma}\label{treefix-runtime-constant}
For a bounded degree tree, treefix sum  takes \(O(n\log n)\) energy and \(O(\log n)\) depth, with high probability.
\end{lemma}

\begin{IEEEproof}
  It takes \(O(\log n)\) repetitions of \textsc{Compact} to contract the tree to a single vertex, with high probability. This follows from the analysis of the Random-mate approach to find an independent set~\cite{pem_lca,MiRe85}. Note that this routine only requires messages between neighboring supervertices. It follows from \Cref{bough analysis constant} that it takes \(O(n \log n)\) energy and $O(\log n)$ depth to contract the tree, with high probability.
  The uncontraction essentially reverses the operations. 
\end{IEEEproof}

\begin{lemma}\label{treefix-runtime-unbounded}
Treefix sum takes \(O(n\log n)\) energy and \(O(\log^2n)\) depth, with high probability.
\end{lemma}
\begin{IEEEproof}
  The proof is analogous to the one for constant degree trees except each round of \textsc{Compact} has depth \(O(\log n)\). 
\end{IEEEproof}

\subsection{Top-down Treefix Sum}\label{sec:top-down-treefix-sum}

Similarly, we can also compute for every vertex $u$ the sum \emph{sum}$'(u)$ of the values along the path from the root to $u$. The only change is in the uncontraction. We maintain the invariant that for every supervertex $u$, \emph{sum}$'(u) = \text{\emph{val}}(u) + A_u$. Initially, $A_r$ is zero for the root supervertex $r$.  When undoing a \textsc{Compress} of a supervertex $u$ with child supervertex $v$, set $A_v=A_u + P_u - P_v$. Then, decrement $P_u$ by $P_v$. When undoing a \textsc{Rake} of a supervertex $u$ with supervertices $v_1, \dotsc, v_i$, set $A_v=A_u + \text{\emph{val}}(u)$ and set $P_u$ to \emph{val}$(u)$. 


\section{Lowest Common Ancestor}\label{sec:lca}
The \emph{lowest common ancestor} (LCA) of two vertices \(u\) and \(v\) is the lowest vertex \(w\), such that both \(u\) and \(v\) are descendants of \(w\). We design an algorithm that processes multiple queries of the form LCA(\(u, v\)) with \(O(n\log n)\) energy and \(O(\log^2 n)\) depth, where each vertex appears in at most a constant number of queries. If this is not the case, the tree can be preprocessed by splitting a vertex with many queries into multiple vertices that form a path and distributing the queries among them.

Our approach is to design an algorithm for LCA that fits into the local messaging framework. Hence, we can benefit from the linear energy bounds of local messaging. We cannot directly use previous approaches~\cite{pem_lca, em_lca, bsp_lca}, as they require non-local messaging.
Using treefix sums, we can answer queries where one vertex is a descendant of the other. For the other types of queries, we introduce a notion of \emph{subtree covers} of the tree. Every vertex is part of at least one and at most $O(\log n)$ subtrees in the cover. The cover guarantees that there is a subtree $S$ such that exactly one of the query vertices is in $S$. This means that the lowest common ancestor is the parent of the root of $S$. Because of this simple structure, we can identify the solution using a series of broadcasts and reductions on the subtrees in the cover.

\subsection{Path Decomposition}
\begin{figure}
\vspace{-1.5em}
\begin{center}

\resizebox{0.5\linewidth}{!}{
\begin{tikzpicture}[step=0.75cm, x=0.75cm, y=0.75cm]

    \node[node,fill=color1] (0) {0};
  
    \node[child1, node distance=17mm] (1) [below left of=0] {$1$};
    \node[node,fill=color1, node distance=17mm] (4) [below right of=0] {$4$};

    \node[child2] (2) [below left of=1] {2};
    \node[child1] (3) [below right of=1] {3};
    \node[child1] (5) [below left of=4] {5};
    \node[node,fill=color1] (6) [below right of=4] {6};
    \node[node,fill=color1] (7) [below left of=6] {7};

    \draw (0) -- (1);
    \draw (0) -- (4);
    \draw (1) -- (2);
    \draw (1) -- (3);
    \draw (4) -- (5);
    \draw (4) -- (6);
    \draw (6) -- (7);

    \node[text=color2] at ($(1) - (1.5, -0.2)$) {$S1$};
    \node[text=color1] at ($(0) + (-1.5, 0.2)$) {$S0$};

    \begin{scope}[on background layer]
      \draw [rounded corners=8mm,fill=color2light] ($(1) + (0,1)$)--($(2) + (-1, -0.5)$)--($(3) + (1, -0.5)$)--cycle;


      \draw [rounded corners=6mm,draw=color1, very thick, dashed] ($(0) + (0,1)$)--($(2) + (-1, -0.3)$)--($(7) + (0.25, -1)$)--($(6) + (1.25,0)$) --cycle;
    \end{scope}

\end{tikzpicture}
}
\end{center}
\vspace{-3.5em}
  \caption{The path decomposition creates a set of $O(\log n)$ layers consisting of disjoint paths. The yellow path $(0, 4, 6, 7)$ is in layer $0$, the green paths $(1,3)$ and $(5)$ are in layer $1$, and the red path $(2)$ is in layer $2$. We get the subtree cover by extending each path to contain all the descendants of the root of the path. Each vertex is labeled with its light-first order. The subtree \(S0\) corresponds to the range \([0, 7]\) and the subtree \(S1\) to \([1, 3]\).}
  \label{fig:ranges}
\end{figure}
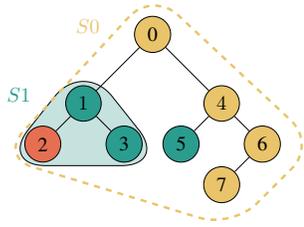

A \emph{path decomposition} $\mathcal{P}$ is a partition of the tree into a set of disjoint paths. Consider a path $P\in \mathcal{P}$ rooted at vertex $v$. If the root-to-$v$ path intersects $i$ other paths in $\mathcal{P}$, then $P$ and $v$ are in the $i$-th layer of $\mathcal{P}$. 
Note that if a vertex $v$ is a child of a vertex $u$ and they are in different paths of the decomposition, then $u$ is in some $i$-th layer and $v$ is in the $(i+1)$-th layer. See Figure \ref{fig:ranges} (top) for an example.

A suitable decomposition, the heavy-light decomposition~\cite{DBLP:conf/esa/Klein98}, can be directly constructed from light-first order: Always connect a vertex with its heaviest child. This is the rightmost child in light-first order. Observe that every time we go down to a child that is not rightmost, the size of the subtree decreases at least by a factor $2$. Hence, connecting vertices via their rightmost children constitutes a path decomposition with $O(\log n)$ layers. We can compute the layer of its path for every vertex with a top-down treefix sum (see \Cref{sec:top-down-treefix-sum}) and an appropriate associative operator. It takes $O(n \log n)$ energy and $O(\log n)$ depth to construct such a decomposition with local messaging.

\subsection{Subtree Cover}

A \emph{subtree cover} is defined given a path decomposition $\mathcal{P}$, as follows. For each path $P$ in the path decomposition $\mathcal{P}$, the subtree cover contains the subtree rooted at the root of that path $P$. We say that a subtree is $i$-th if it is rooted at a vertex on the $i$-th layer. Note that all $i$-th subtrees are pairwise vertex disjoint, but subtrees from different levels overlap. See Figure~\ref{fig:ranges} for an example subtree cover.

Next, we show the main structural lemma regarding subtree covers. 
We say that a vertex \(w\) is a parent of a subtree \(S\) rooted at vertex \(v\) if \(w\) is a parent of \(v\).

\begin{corollary}\label{LCA bough}
Consider a subtree cover and let LCA(\(u, v\)) = \(w\). Then, either \(w \in \{u, v\}\) or \(w\) is a parent of a subtree in the cover that contains exactly one of those vertices.
\end{corollary}
\begin{IEEEproof}
We assume that \(w \notin \{u, v\}\) and show that \(w\) is a parent of a subtree in the cover that contains one of the vertices. Let $\mathcal{P}$ be the path decomposition that defined the subtree cover. 
Because  \(w \notin \{u, v\}\), there must be two children $x_u$ and $x_v$ of $w$ that are ancestors of $u$ and $v$, respectively. Only one of those two vertices can be in the same path of the path decomposition $\mathcal{P}$ as $w$. The other vertex must therefore be the root of a subtree in the cover. 
\end{IEEEproof}

\subsection{Local Messaging LCA Algorithm}

We assume that the input tree is stored in energy-bound light-first order. For each query LCA(\(u, v\)), we assume the input is stored in both \(u\) and \(v\). The result of the query is stored in one of the two vertices. 

Consider a \emph{range} \([a, b]\) for integers \(0 \leq a \leq b \leq n\). We say a vertex \(v \in [a, b]\) if \(v\) is stored in the \(i\)-th processor in the message bound curve order for some $i$ with \(a \leq i \leq b\). Each subtree $S$ rooted at $v$ corresponds to exactly one contiguous range \(r(v)=[a, b]\) that contains all its descendants. We can test if a vertex is in a subtree given the range of its root. See Figure \ref{fig:ranges} for an example.

Our local messaging LCA algorithm has four main steps:
\begin{enumerate}[leftmargin=13pt]
  \item Let each vertex have value 1 and run the treefix sum algorithm such that each vertex \(u\) contains a value \(sum(u)\), which is the size of the subtree rooted at the vertex $u$. The range of a subtree rooted at a vertex $u$ stored in $i$-th position in light-first order is now \(r(u) = [i, i + sum(u) - 1]\). Then, answer all queries where LCA(\(u, v\)) \(\in \{u, v\}\). If \(v \in r(u)\) then \(u\) answers that query; otherwise, \(v\) answers it.
  \item Every vertex \(w\) does a local broadcast to send the range of its subtree \(r(w)\) to its children.
  \item Compute the path decomposition using a top-down treefix sum.
  \item Let \(\mathcal{B}\) be the subtree cover. For each layer $i$ in increasing order, consider all subtrees $S$ in that layer. Let $w$ be the parent of the subtree $S$ and let $x$ be its root.
   \begin{enumerate}
              \item Broadcast \(r(w) \backslash r(x)\) within $S$, where \(r(w)\backslash r(x)\) is the range of \(w\) excluding the range of \(x\).
              \item If \(u\) is in the subtree $S$, we know that \(LCA(u, v) = w\) if \(v \in r(w) \backslash r(x)\). We answer those queries.
    \end{enumerate} 
  Perform a synchronization barrier before proceeding to the next layer. This can be performed by doing an all-reduce on the compute grid, where a processor starts the all-reduce once it hits the barrier.
\end{enumerate}
The correctness of the algorithm follows directly from Corollary \ref{LCA bough}. If \(LCA(u, v) = w \notin \{u, v\}\), at some point exactly one of the vertices will be part of a subtree which is a child of \(w\). The query will therefore always have a correct answer.
\begin{lemma}\label{broadcast_range}
  It takes \(O(b - a)\) energy and \(O(\log(b - a))\) depth to broadcast a message on the range \([a, b]\) with \(a < b\) when the tree is stored in energy-bound light-first order.
\end{lemma}

\begin{IEEEproof}
We build a virtual broadcast tree on the processors in the range, such that the tree is a complete binary tree in energy-bound light-first order. Because the range is contiguous, such a tree can be constructed top-down using only the indexes of the range. The statement now follow from light first order being message bound. 
\end{IEEEproof}

\begin{theorem}
  The local messaging LCA algorithm takes \(O(n\log n)\) energy and $O(\log ^2 n)$ depth with high probability. 
\end{theorem}
\begin{IEEEproof}
With high probability, the first three steps take $O(n\log n)$ energy and $O(\log ^2 n)$ depth, by Lemma~\ref{treefix-runtime-unbounded}. 
Note that every subtree in the cover corresponds to a range and that subtrees from the same level are disjoint. Hence, the sum of their ranges is at most \(n\). By Lemma~\ref{broadcast_range}, we can do the broadcast within all the \(i\)-th subtrees in \(O(n)\) energy. The barrier takes $O(n)$ energy and $O(\log n)$ depth using an all-reduce~\cite{SpatialModel}. We conclude that the fourth step takes \(O(n \log n)\) energy overall. For each level, the depth is $O(\log n)$ for the broadcasts and the barrier. Hence, the depth is $O(\log ^ 2 n)$. 
\end{IEEEproof}

This near-linear energy and low depth algorithm constitutes a significant improvement on the na\"ive PRAM simulation, which would take $\Omega(n^{\frac{3}{2}})$ energy~\cite{SpatialModel}.

\section{Conclusion}\label{sec:conclusion}

This paper presents spatial tree algorithms that minimize communication distances between vertices on a 2D computational grid through a twofold framework:
\begin{enumerate}
\item \emph{Data Layout}: Our strategy mapped high-dimensional tree structures onto a localized linear layout. This was subsequently lifted to the 2D grid using space-filling curves. This layout reduces the average distance between neighboring vertices to a \emph{constant}, optimizing communication.
\item \emph{Logical Operations}: By separating logical operations from the layout, we could leverage existing strategies from the PRAM environment. This separation allowed us to maintain low depth and work while also achieving low energy. We addressed two foundational tree problems: treefix sum and lowest common ancestors. For both approaches, our algorithms exhibited near-linear energy and poly-logarithmic depth.
\end{enumerate}
This framework could be fruitful for optimizing other sparse workloads, including sparse matrix-vector multiplication and graph clustering.
Future exploration of layouts supporting dynamic updates may enhance the real-time adaptability of our framework. Not only could this address current limitations that require layouts to be precomputed, but it could also pave the way for more dynamic and versatile applications.

In conclusion, this research constitutes a novel approach to deploying specialized architectures, such as the Wafer-Scale engine~\cite{Cerebras} and CGRAs~\cite{DBLP:conf/fpga/Vissers19}. Our work addresses irregular sparse workloads and unlocks new possibilities in large-scale data science applications.
\section*{Acknowledgment}
This project received funding from the European Research Council (Project PSAP, No. 101002047). 
This project was supported by the ETH Future Computing Laboratory (EFCL), financed by a donation from Huawei Technologies. This project received funding from the European Research Council under the European Union's Horizon 2020 programme GLACIATION, No. 101070141. T.B.N. was supported by the Swiss National Science Foundation (Ambizione Project No. 185778). The language in the abstract, \Cref{sec:introduction}, and \Cref{sec:conclusion} has been enhanced with a large language model~\cite{chatgpt}.

\IEEEtriggeratref{47}

\bibliographystyle{IEEETrans}
\bibliography{citations}

\clearpage

\appendices

\section{Proof of \Cref{minimized}}\label{sec:appendix}

We begin with the following inequality:

\begin{lemma}\label{inequality}
  Let \(a, b\) be positive constants with \(\frac{b}{2} \leq a \leq b\), then \(\forall 0 \leq 2x \leq y\) we have \(b\sqrt{y} \leq a\sqrt{x} + b\sqrt{y - x}\).
\end{lemma}
\begin{IEEEproof}
  Let \(f(x, y) = b\sqrt{y}\) and \(g(x, y) = a\sqrt{x} + b\sqrt{y - x}\). We show that \(f^2(x, y) - g^2(x, y) \leq 0\), which implies the result. 
  If \(x = 0\), then we have equality trivially. For \(x > 0\), we have:
  \begin{align*} \label{eq2}
      &f^2(x, y) - g^2(x, y)  \\
      = \ &b^2y - a^2x - 2ab\sqrt{x(y - x)} - b^2y + b^2x\\
      \leq &\left(b^2 - \left(\frac{b}{2}\right)^2 - 2\frac{b}{2}b\right) x\\
      \leq &-\frac{b^2}{4}\cdot x \leq 0 \enspace . 
  \end{align*}
  We conclude using that $f(x, y)$ and $g(x, y)$ are nonnegative.
\end{IEEEproof}

We continue with the proof of \Cref{minimized}.
\begin{IEEEproof}[Proof of \Cref{minimized}]
  Let us assume we start with \(s({c_\Delta}) = n\) and the other variables set to 0, i.e., our value is \(2\Delta \cdot c\sqrt{n}\). We can create any combination that fulfils the constraints by applying \(\Delta - 1\) transformations. For \(i = 1, ..., \Delta - 1\), we increment \(s({c_i})\) by some value \(x_i\) and subtract that value from \(s({c_d})\). The problem is equivalent to the one in Lemma \ref{inequality}: Let \(y_i\) be the value of \(s({c_\Delta})\) before the \(i\)-th transformation. In our case we are comparing \(b\sqrt{y_i}\) with \(a\sqrt{x_i} + b\sqrt{y_i - x_i}\), where \(\frac{b}{2} = \Delta \cdot c \leq a = (\Delta + i) \cdot c \leq 2\Delta \cdot c = b\). Since \(s({c_\Delta}) \geq s({c_i})\), we have \(0 \leq 2x_i \leq y_i\). Hence, no such transformation decreases the sum; the function is minimized for \(s({c_\Delta}) = n\).
\end{IEEEproof}

\end{document}